%% file: reportarxiv.tex
\documentclass[runningheads,a4paper]{llncs}

\usepackage{t1enc}
\usepackage[utf8]{inputenc}

\usepackage{hyperref} 
\usepackage{amssymb} 
\usepackage{amsmath} 

\usepackage{listings}

\input{macros} 

\usepackage{bcprules} 

\usepackage{stmaryrd}
\newcommand{\comp}{\fatsemi}

\usepackage{url}

\newcommand{\muSPARK}{$\mu$SPARK}

\begin{document}

\title{Borrowing Safe Pointers from Rust in SPARK}


\author{Georges-Axel Jaloyan\inst{1}%
\thanks{This work is partly supported by the Joint Laboratory ProofInUse
(ANR-13-LAB3-0007) and project VECOLIB (ANR-14-CE28-0018) of the French
National Research Agency (ANR).}
\and Yannick Moy\inst{2}
\and Andrei Paskevich\inst{3,4}}

\authorrunning{Georges-Axel Jaloyan et al.} 


\institute{\'Ecole Normale Supérieure, \\
PSL Research University, Paris, France 
\and
AdaCore, Paris, France 
\and
Laboratoire de Recherche en Informatique, \\
Universit\'e Paris-Sud, CNRS, Orsay, F-91405
\and
Inria Saclay, Universit\'e Paris Saclay, Palaiseau, F-91120
}

\maketitle

\begin{abstract}
In the field of deductive software verification, programs with
pointers present a major challenge due to pointer aliasing.
In this paper, we introduce pointers to SPARK, a well-defined
subset of the Ada language, intended for formal verification
of mission-critical software.
Our solution uses a permission-based static alias analysis
method inspired by
Rust's \emph{borrow-checker} and \emph{affine types}, and
enforces the \textsl{Concurrent Read, Exclusive Write} policy.
This analysis has been implemented in the GNAT Ada compiler
and tested against a number of challenging examples.
In the paper, we give a formal presentation of the analysis
rules for a miniature version of SPARK and prove their soundness.
We discuss the implementation and compare our solution with Rust.
\end{abstract}


\section{Introduction}

SPARK~\cite{mccormick_chapin_2015} is a subset of the Ada programming language
targeted at safety- and security-critical applications. SPARK restrictions
ensure that the behavior of a SPARK program is unambiguously defined, and
simple enough that formal verification tools can perform an automatic diagnosis
of conformance between a program specification and its implementation.

As a consequence of SPARK's focus on automation and usability, it forbids
the use of programming language features that either prevent
automatic proof, or make it possible only at the expense of extensive user
effort in annotating the program. The lack of support for pointers in SPARK is
the main example of this choice. While it is possible to exclude parts of the
programs that manipulate pointers from analysis, it would be preferable to
support pointers when their use does not prevent formal verification.

Among the various problems related to the use of pointers in the context of
formal program verification, the most difficult problem is the possibility that
two names refer to overlapping memory locations, a.k.a.~aliasing.  Formal
verification platforms that support pointer aliasing like
Frama-C~\cite{Kirchner2015} require users to annotate programs to specify when
pointers are not aliased.  This can take the form of inequalities between
pointers when a typed memory model is used, or the form of separation
predicates between memory zones when an untyped memory model is used. In both
cases, the annotation burden is acceptable for leaf functions which manipulate
single-level pointers, and quickly becomes overwhelming for functions that
manipulate pointer-rich data structures. In parallel to the increased cost of
annotations, the benefits of automation decrease, as automatic provers have
difficulties reasoning explicitly with these inequalities and separation
predicates.

Programs often rely on non-aliasing in general for correctness, when such
aliasing would introduce interferences between two unrelated names. We call
such aliasing \emph{potentially harmful} when a memory location modified
through one name could be read through another name, within the scope of a
verification condition. Otherwise, the aliasing is \emph{benign}, when the
memory location is only read through both names. A reasonable restriction for
formal program verification is thus to forbid potentially harmful aliasing of
names. The difficulty is then to guarantee the absence of potentially harmful
aliasing. The following code shows an example where we want analysis to be able
to rely on the non-aliasing of parameters \verb|X| and \verb|Y| to prove the
postcondition of the procedure \verb|Assign_Incr|:
\begin{lstlisting}[style=spark]
 procedure Assign_Incr (X, Y : in out Integer_Pointer)
   with Post => Y.all = X.all + 1
 is
 begin
   Y.all := X.all + 1;
 end Assign_Incr;
\end{lstlisting}

In this work, we present the first step for the inclusion of pointers in the
Ada language subset supported in SPARK. As our main contribution, we show that
it is possible to borrow and adapt the ideas underlying the safe support for
pointers in permission-based languages like Rust, to safely restrict the use of
pointers in usual imperative languages like Ada. This adaptation is based
on a possible division of work between a permission-based anti-aliasing
analysis, lifetime management by typing, and the use of a formal verification
platform for checking non-nullity of accessed pointers. For example, these
rules prevent aliasing between parameters \verb|X| and \verb|Y| in the code of
procedure \verb|Assign_Incr| above, which makes it possible to treat pointers
in proof like records with a single field corresponding to the type of the
object pointed to. Thus, the verification condition corresponding to the
postcondition of procedure \verb|Assign_Incr| has a form (using
\verb|get|/\verb|set| to access the field \verb|all| of variables
\verb|X| and \verb|Y|) that can readily be proved by automatic provers:
\begin{verbatim}
 hypothesis: Y' = set(Y, all, get(X, all) + 1)
 goal:       get(Y', all) = get(X, all) + 1
\end{verbatim}

In Section~\ref{sec:pointer-analysis}, we present a formalization of the
non-aliasing rules enforced by the permission-based analysis, including proof
of non-aliasing guarantees. In Section~\ref{sec:experimental}, we describe a
concrete implementation of the analysis inside the open-source GNAT compiler
for Ada which is part of GCC. We survey related works in
Section~\ref{sec:related}, in particular with respect to Rust.

    \section{Alias Analysis}
    \label{sec:pointer-analysis}

\newcommand{\RW}{\mathsf{RW}}
\newcommand{\RO}{\mathsf{R}}
\newcommand{\WO}{\mathsf{W}}
\newcommand{\NO}{\mathsf{NO}}

In Ada/SPARK code, the access to memory areas is given through
\emph{paths} that start with an identifier (a variable name)
and follow through record fields, array indices, or through
a special field \texttt{all}, which corresponds to pointer
dereferencing. In this paper, we only consider record
and pointer types, and discuss the treatment of arrays
in Section~\ref{sec:completeSPARK}.

As an example, we use the following Ada type,
describing singly linked lists where each node carries a
boolean flag and a pointer to a shared integer value.
\begin{lstlisting}[style=spark]
 type List is record
   Flag : Boolean;
   Key  : access Integer;
   Next : access List;
 end record;
\end{lstlisting}

Given a variable \texttt{A} \texttt{:} \texttt{List}, the paths
\texttt{A.Flag}, \texttt{A.Key.all}, \texttt{A.Next.all.Key}
are valid and their respective types are \texttt{Boolean},
\texttt{Integer}, and \texttt{access Integer} (a pointer
to an \texttt{Integer}). The important difference between pointers
and records in Ada is that---similarly to C---assignment
of a record copies the values of fields, whereas assignment
of a pointer only copies the address and creates an alias.

The alias analysis procedure runs after the type checking.
The idea is to associate one of the four permissions---%
$\RW$, $\RO$, $\WO$ or $\NO$---to each possible path (starting
from the available variables) at each sequence point in
the program.

The \emph{read-only} permission $\RO$ allows us to read any
value accessible from the path: use it in a computation,
or pass it as an \texttt{in} parameter in a procedure call.
As a consequence, if a given path has the $\RO$ permission,
then each valid extension of this path also has it.

The \emph{write-only} permission $\WO$ allows us to modify memory
occupied by the value: use it on the left-hand side in an assignment
or pass it as an \texttt{out} parameter in a procedure call. For example,
having a write permission for a path of type \texttt{List} allows
us to modify the \texttt{Flag} field or to change the addresses
stored in the pointer fields \texttt{Key} and \texttt{Next}.
However, this does not necessarily give us the permission
to modify memory accessible from those pointers.
Indeed, to dereference a pointer, we must read the address
stored in it, which requires the read permission.
Thus, the $\WO$ permission only propagates to path extensions
that do not dereference pointers, i.e.,~do not contain
additional \texttt{all} fields.

The \emph{read-write} permission $\RW$ combines the properties
of the $\RO$ and $\WO$ permissions and grants the full ownership
of the path and every value accessible from it. In
particular, the $\RW$ permission propagates to all valid
path extensions including those that dereference pointers.
The $\RW$ permission is required to pass a value as an
\texttt{in-out} parameter in a procedure call.

Execution of program statements changes permissions. For example,
allocating a new non-initialised memory area assigns
the $\WO$ permission to every value stored in this area.
Thus, after the statement \texttt{P := new List}, the paths
\texttt{P}, \texttt{P.all}, \texttt{P.all.Flag},
\texttt{P.all.Key}, and \texttt{P.all.Next}
have permission $\WO$. All strict extensions of \texttt{P.all.Key}
and \texttt{P.all.Next} receive permission $\NO$, that is, no
permission. Indeed, since the pointers \texttt{P.all.Key} and
\texttt{P.all.Next} are not initialised, neither reads nor writes
under them make any sense. Another simple example of permission
change is the procedure call: all \texttt{out} parameters must
be assigned by the callee and receive the $\RW$ permission
after the call.

The assignment statement is more complicated and several
cases must be considered. If we assign a value that does
not contain pointers (say, an integer or a pointer-free
record), the whole value is copied into the left-hand side,
and we only need to check that we have the appropriate
permissions: $\WO$ or $\RW$ for the left-hand side and $\RO$ or $\RW$
for the right-hand side. However, whenever we copy a pointer,
an alias is created. We want to make the left-hand side
the new full owner of the value (i.e.,~give it the $\RW$
permission), and therefore, after the permission checks,
we must revoke the permissions from the right-hand side,
to avoid potentially harmful aliasing. The permission checks
are also slightly different in this case, as we require the
right-hand side to have the $\RW$ permission in order to
move it to the left-hand side.

Let us now consider several simple programs and see
how the permission checks allow us to detect
potentially harmful aliasing.

\begin{figure}[t]
\begin{minipage}[t]{0.41\textwidth}
\begin{lstlisting}[style=spark]
procedure P1
  (A,B: in out List) is
begin
  A := B;
  B.Flag := True;
  B.Key.all := 42;
end P1;
\end{lstlisting}
\end{minipage}
\begin{minipage}[t]{0.585\textwidth}
\begin{lstlisting}[style=spark]
procedure P2
  (A,B: in out Integer_Pointer) is
begin
  while B.all > 0 loop
    A.all := A.all + 1;
    B.all := B.all - 1;
    A := B;
  end loop;
end P2;
\end{lstlisting}
\end{minipage}
\caption{Examples of potentially harmful aliasing.}
\label{fig:P1P2}
\end{figure}

Procedure \texttt{P1} in Fig.~\ref{fig:P1P2} receives two \texttt{in-out}
parameters \texttt{A} and \texttt{B} of type \texttt{List}.
At the start of the procedure, all \texttt{in-out}
parameters assume permission $\RW$. In particular, this implies
that each \texttt{in-out} parameter is separated from all other
parameters (in fact, only the \texttt{in} parameters may alias
each other). The first assignment copies the structure \texttt{B}
into \texttt{A}. Thus, the paths
\texttt{A.Flag}, \texttt{A.Key}, and \texttt{A.Next}
are separated, respectively, from
\texttt{B.Flag}, \texttt{B.Key}, and \texttt{B.Next}.
However, the paths \texttt{A.Key.all} and \texttt{B.Key.all}
are aliased, and \texttt{A.Next.all} and \texttt{B.Next.all}
are aliased as well.

The first assignment does not change the
permissions of \texttt{A} and its extensions: they retain the
$\RW$ permission and keep the full ownership of their respective
memory areas, even if the areas themselves have changed.
The paths under \texttt{B}, however, must relinquish (some of) their
permissions. The paths \texttt{B.Key.all} and \texttt{B.Next.all}
as well as all their extensions get the $\NO$ permission, that
is, lose both read and write permissions. This is necessary, as
the ownership over their memory areas is transferred to the
corresponding
paths under \texttt{A}. The paths \texttt{B}, \texttt{B.Key},
and \texttt{B.Next} lose the read permission but keep
the write-only $\WO$ permission. Indeed, we forbid reading
from memory that can be altered through a concurrent path.
However, it is allowed to ``redirect'' the pointers
\texttt{B.Key} and \texttt{B.Next}, either by assigning
those fields directly or by copying some different record
into \texttt{B}. The field \texttt{B.Flag} is not aliased,
nor has it aliased extensions, and thus retains the initial
$\RW$ permission. This $\RW$ permission allows us to perform
the assignment \texttt{B.Flag := True} on the next line.

The third assignment, however, is now illegal, since
\texttt{B.Key.all} does not have the write permission anymore.
What is more, at the end of the procedure the \texttt{in-out}
parameters \texttt{A} and \texttt{B} are not separated.
This is forbidden, as the caller assumes that all \texttt{out}
and \texttt{in-out} parameters are separated after the call
just as they were before.

Procedure \texttt{P2} in Fig.~\ref{fig:P1P2} receives two pointers
\texttt{A} and \texttt{B}, and manipulates them inside a while loop.
Since the permissions are assigned statically, we must ensure that
at the end of a single iteration, we did not lose the permissions
necessary for the next iteration. This requirement is violated in
the example: after the last assignment \texttt{A := B}, the path
\texttt{B} receives permission $\WO$ and the path \texttt{B.all},
permission $\NO$, as \texttt{B.all} is now an alias of \texttt{A.all}.
The new permissions for \texttt{B} and \texttt{B.all} are thus
weaker than the original ones ($\RW$ for both), and the procedure
is rejected. Should it be accepted, we would have conflicting
memory modifications from two aliased paths at the beginning
of the next iteration.

    \subsection{{\muSPARK} language}
    \label{ssec:muspark}

For the purposes of formal presentation, we introduce {\muSPARK},
a small subset of SPARK featuring pointers, records, loops,
and procedure calls. We present the syntax
and semantics of {\muSPARK}, and define the rules for static
analysis of alias safety.

The data types of \muSPARK{} are as follows:
$$
\begin{array}{rclcl}
type & \:::=\; & \texttt{Integer} \:\:|\:\: \texttt{Real} \:\:|\:\:
\texttt{Boolean} & \qquad & \text{scalar type} \\
&|& \texttt{access} \:\: type && \text{access type (pointer)} \\
&|& ident && \text{record type}
\end{array}
$$

Every {\muSPARK} program starts with a list of record type declarations:
$$
\begin{array}{rcl}
record & \:::=\; &
\texttt{type}\:\: ident \:\: \texttt{is} \:\: \texttt{record} \:\:
field^\star_\texttt{;} \:\: \texttt{end} \\[1ex]
field & \:::=\; & ident \:\: \texttt{:} \:\: type
\end{array}
$$

We require all field names to be distinct. The field types must
not refer to the record types declared later in the list.
Recursive record types are allowed: a field of a record type $R$
can contain pointers to $R$ (written \texttt{access} $R$).
We discuss the handling of array types in Section~\ref{sec:completeSPARK}.

The syntax of {\muSPARK} statements is defined by the following rules:
$$
\begin{array}{rclcl}
path & \:::=\; & ident & \qquad & \text{variable} \\
&|& path \,\texttt{.}\, ident && \text{record field} \\
&|& path \,\texttt{.}\, \texttt{all} && \text{pointer dereference}
\\[2ex]
expr & \:::=\; & path && \text{l-value} \\
&|& \texttt{42} \;|\; \texttt{3.14} \;|\; \texttt{True} \;|\;
\texttt{False} \;|\; \ldots & \qquad & \text{scalar value} \\
&|& expr
\:\:(\,\texttt{+}\,|\,\texttt{-}\,|\,\texttt{<}\,|\,\texttt{=}\,|\,\ldots\,)\:\: expr
&& \text{binary operator} \\
&|& path \texttt{'Access} && \text{address of an l-value} \\
&|& \texttt{null} && \text{null pointer}
\\[2ex]
stmt & \:\:::=\; & path \:\:\texttt{:=}\;\; expr  && \text{assignment} \\
&|& path \:\:\texttt{:=}\;\; \texttt{new} \:\: type && \text{allocation} \\
&|& \texttt{if} \:\:expr\:\: \texttt{then}\:\: stmt^{\star}_\texttt{;} \:\: \texttt{else} \:\:
  stmt^{\star}_\texttt{;} \:\: \texttt{end} && \text{conditional} \\
&|& \texttt{while} \:\:expr\:\: \texttt{loop}\:\: stmt^{\star}_\texttt{;} \:\: \texttt{end}
  && \text{``while'' loop} \\
&|& ident \:\:\texttt{(}\:\: { expr }^{\star}_\texttt{,} \:\:\texttt{)}
  && \text{procedure call}
\end{array}
$$

Following the record type declarations,
a {\muSPARK} program contains
a set of mutually recursive procedure declarations:
$$
\begin{array}{rcl}
procedure & \:::=\; &
\texttt{procedure} \:\: ident \:\:
\texttt{(} \:\: param^{\star}_\texttt{;} \:\: \texttt{)} \:\:
\texttt{is} \:\: local^{\star}_\texttt{;} \:\:
\texttt{begin}\:\: stmt^{\star}_\texttt{;} \:\: \texttt{end} \\[1ex]
param & \:::=\; & ident \:\:\texttt{:}\:\:
(\, \texttt{in}\:|\: \texttt{in-out} \:|\: \texttt{out}\,) \:\: type \\[1ex]
local & \:::=\; & ident \:\:\texttt{:}\:\: type
\end{array}
$$
We require all formal parameters and
local variables in a procedure to have distinct names.
A procedure call can only pass left-values (i.e.,~paths)
for \texttt{in-out} and \texttt{out} parameters.
The execution starts from a procedure named
\texttt{Main} with the empty parameter list.

The type system for {\muSPARK} is rather standard and
we do not present it here in full.
We assume that binary operators only operate on scalar types.
The null pointer can have any pointer type \texttt{access $\tau$}.
The dereference operator \texttt{.all} converts a pointer
type \texttt{access $\tau$} to $\tau$. The access operator
\texttt{'Access} applied to an \mbox{l-value} of type $\tau$
returns the corresponding pointer type \texttt{access $\tau$}.
The allocation statement \texttt{$p$ := new $\tau$} requires
the path $p$ to have type \texttt{access $\tau$}.
In what follows, we only consider well-typed {\muSPARK} programs.

On the semantic level, we need to distinguish the units
of allocation, such as whole records, from the units of
access, such as individual record fields. We use the term
\emph{location} to refer to the memory area occupied by
an allocated value. We treat locations as elements of an
abstract infinite set, and denote them with letter $\ell$.
We use the term \emph{address} to designate either
a location, denoted $\ell$, or a specific component
inside the location of a record, denoted $\ell.f.g$,
where $f$ and $g$ are field names (assuming that
at $\ell$ we have a record whose field $f$
is itself a record with a field~$g$).
A \emph{value} is either a scalar, an address,
a null pointer or a record, that is, a finite
mapping from field names to values.

\newcommand{\lval}[1]{\langle #1 \rangle}
\newcommand{\eval}[1]{\llbracket #1 \rrbracket}

A {\muSPARK} program is executed in the context defined
by a \emph{binding} $\Upsilon$ that maps variable names
to addresses and a \emph{store} $\Sigma$ that maps locations
to values. By a slight abuse of notation, we apply $\Sigma$
to arbitrary addresses, so that $\Sigma(\ell.f)$ is
$\Sigma(\ell)(f)$, the value of the field $f$ of
the record value stored in $\Sigma$ at $\ell$. Similarly,
we write $\Sigma[\ell.f \mapsto v]$ to denote
an update of a single field in a record, that is,
$\Sigma[\ell \mapsto \Sigma(\ell)[f \mapsto v]]$.

We use big-step operational semantics and write
$\Upsilon \cdot \Sigma \cdot s \Downarrow \Sigma'$
to denote that {\muSPARK} statement $s$, when evaluated
under binding $\Upsilon$ and store $\Sigma$, terminates
with the state of the store $\Sigma'$.
We extend this notation to sequences of statements
$\bar{s}$ in an obvious way.
In this paper, we do not consider diverging statements.

The evaluation of expressions is effect-free and is denoted
$\eval{e}^\Upsilon_\Sigma$. We also need
to evaluate l-values to the corresponding addresses
in the store, written $\lval{p}^\Upsilon_\Sigma$, where
$p$ is the evaluated path. Illicit operations, such as
dereferencing a null pointer, cannot be evaluated and
stall the execution (\emph{blocking semantics}).
In the formal rules below, $c$ stands for
a scalar constant and $\odot$, for a binary operator:
\begin{align*}
\lval{x}^\Upsilon_\Sigma &= \Upsilon(x) &
\lval{p.f}^\Upsilon_\Sigma &= \lval{p}^\Upsilon_\Sigma.f &
\lval{p.\texttt{all}}^\Upsilon_\Sigma &= \eval{p}^\Upsilon_\Sigma
\\[1ex]
\eval{c}^\Upsilon_\Sigma &= c &
\eval{p}^\Upsilon_\Sigma &= \Sigma(\lval{p}^\Upsilon_\Sigma) &
\eval{p\texttt{'Access}}^\Upsilon_\Sigma &= \lval{p}^\Upsilon_\Sigma
\\[1ex]
&&
\eval{e_1 \odot e_2}^\Upsilon_\Sigma &=
  \eval{e_1}^\Upsilon_\Sigma \odot \eval{e_2}^\Upsilon_\Sigma &
\eval{\texttt{null}}^\Upsilon_\Sigma &= \texttt{null}
\end{align*}

Allocation adds a fresh address
to the store, mapping it to a default value for
the corresponding type: 0 for \texttt{Integer},
\texttt{False} for \texttt{Boolean},
\texttt{null} for the access types, and for
the record types, a record value where each field
has the default value. Notice that since pointers
are initialised to \texttt{null}, there is no
deep allocation. We write $\Box_\tau$ to denote
the default value of type $\tau$.

The evaluation rules are given in Figure~\ref{fig:sem}.
In the (\textsc{E-call}) rule, we evaluate the procedure
body in the dedicated context $\Upsilon_P \cdot \Sigma_P$.
This context binds the \texttt{in} parameters 
to fresh locations containing the values
of the respective expression arguments, 
binds the \texttt{in-out} and \texttt{out}
parameters 
to the addresses of the respective l-value arguments,
and allocates memory for the local variables. 
For simplicity, we do not reclaim memory on return
from a procedure call, and thus avoid dangling
pointers. In Ada and SPARK, this issue is handled
separately, using scope-based memory pools, and
does not need to be addressed by our analysis procedure.

\begin{figure}[t]
\infrule[E-assign]{\eval{e}^\Upsilon_\Sigma = v}
  {\Upsilon \cdot \Sigma \cdot p \,\texttt{:=}\, e \:\Downarrow\:
   \Sigma[\,\lval{p}^\Upsilon_\Sigma \mapsto v\,]}

\medskip
\infrule[E-alloc]{\ell \not\in dom \:\Sigma}
  {\Upsilon \cdot \Sigma \cdot p \,\texttt{:=}\, \texttt{new}\,\tau \:\Downarrow\:
   \Sigma[\,\lval{p}^\Upsilon_\Sigma \mapsto \ell,\, \ell \mapsto \Box_\tau\,]}

\medskip
\infrule[E-ifTrue]{\eval{e}^\Upsilon_\Sigma = \texttt{True}
  \qquad \Upsilon \cdot \Sigma \cdot \bar{s}_1 \:\Downarrow\: \Sigma'}
  {\Upsilon \cdot \Sigma \cdot \text{\texttt{if $e$ then $\bar{s}_1$ else $\bar{s}_2$}}
  \:\Downarrow\: \Sigma'}

\medskip
\infrule[E-ifFalse]{\eval{e}^\Upsilon_\Sigma = \texttt{False}
  \qquad \Upsilon \cdot \Sigma \cdot \bar{s}_2 \:\Downarrow\: \Sigma'}
  {\Upsilon \cdot \Sigma \cdot \text{\texttt{if $e$ then $\bar{s}_1$ else $\bar{s}_2$}}
  \:\Downarrow\: \Sigma'}

\medskip
\infrule[E-whileTrue]{\eval{e}^\Upsilon_\Sigma = \texttt{True}
  \qquad \Upsilon \cdot \Sigma \cdot
    (\text{\texttt{$\bar{s}$ ; while $e$ loop $\bar{s}$ end}})
  \:\Downarrow\: \Sigma'}
  {\Upsilon \cdot \Sigma \cdot \text{\texttt{while $e$ loop $\bar{s}$ end}}
  \:\Downarrow\: \Sigma'}

\medskip
\infrule[E-whileFalse]{\eval{e}^\Upsilon_\Sigma = \texttt{False}}
  {\Upsilon \cdot \Sigma \cdot \text{\texttt{while $e$ loop $\bar{s}$ end}}
  \:\Downarrow\: \Sigma}

\medskip
\infrule[E-call]{
\texttt{procedure} \:P\:
\texttt{(}\,
a_1 \,\texttt{:}\,\texttt{in}\: \tau_{a_1} \texttt{;} \ldots \texttt{;}\,
b_1 \,\texttt{:}\,\texttt{in-out}\: \tau_{b_1} \texttt{;} \ldots \texttt{;}\,
c_1 \,\texttt{:}\,\texttt{out}\: \tau_{c_1} \texttt{;} \ldots \texttt{)} \\
\text{~~~~~~~~~~~~~~~~~~}
\texttt{is}\; d_1 \,\texttt{:}\, \tau_{d_1} \texttt{;} \ldots \:
\texttt{begin} \; \bar{s} \; \texttt{end}
\text{~~is declared in the program}\\[1ex]
\ell_{a_1},\ldots, \ell_{d_1},\ldots \not\in dom \:\Sigma
\qquad\qquad
\eval{e_{a_1}}^\Upsilon_\Sigma ,\ldots = v_{a_1},\ldots\\[1ex]
\Upsilon_P = [\,a_1 \mapsto \ell_{a_1}, \ldots,
b_1 \mapsto \lval{p_{b_1}}^\Upsilon_\Sigma, \ldots,
c_1 \mapsto \lval{q_{c_1}}^\Upsilon_\Sigma, \ldots,
d_1 \mapsto \ell_{d_1}, \ldots\,]\\[1ex]
\Sigma_P = \Sigma[\,\ell_{a_1} \mapsto v_{a_1}, \ldots,
  \ell_{d_1} \mapsto \Box_{\tau_{d_1}}, \ldots\,] \qquad\qquad
\Upsilon_P \cdot \Sigma_P \cdot \bar{s} \:\Downarrow\: \Sigma'
}{
\Upsilon \cdot \Sigma \cdot P(e_{a_1},\ldots,
p_{b_1},\ldots,q_{c_1},\ldots) \Downarrow \Sigma'}

\caption{Semantics of {\muSPARK} (terminating statements).}\label{fig:sem}
\end{figure}

\subsection{Access policies, transformers, and alias safety rules}

We denote paths with letters $p$ and $q$.
We write $p \sqsubset q$ to denote that
$p$ is a strict \emph{prefix} of $q$ or, equivalently,
$q$ is a strict \emph{extension} of $p$. In what follows,
we always mean strict prefixes and extensions, unless
explicitly said otherwise.

In the typing context of a given procedure,
a well-typed path is said to be \emph{deep} if it has
an extension of an access type, otherwise it is called
\emph{shallow}.
We extend these notions to types:
a type $\tau$ is deep (resp.~shallow) if
and only if a $\tau$-typed path is deep (resp.~shallow).
In other words, a path or a type is deep if a pointer
can be reached from it, and shallow otherwise.
For example, the \texttt{List} type in
Section~\ref{sec:pointer-analysis} is a deep type,
and so is \texttt{access} \texttt{Integer}, whereas
any scalar type
or any record with scalar fields only is shallow.

An extension $q$ of a path $p$ is called a
\emph{near extension} if it has as many pointer
dereferences as $p$, otherwise it is a
\emph{far extension}. For instance, given a variable
\texttt{A} of type \texttt{List}, the paths \texttt{A.Flag},
\texttt{A.Key}, and \texttt{A.Next} are the near extensions
of \texttt{A}, whereas \texttt{A.Key.all}, \texttt{A.Next.all},
and their extensions are far extensions, since they all create
an additional pointer dereference by passing through \texttt{all}.

We say that \emph{sequence points}
are the program points before or after a given statement.
For each sequence point in a given \muSPARK{} program, we
statically compute an \emph{access policy}: a partial function
that maps each well-typed path to one of the four \emph{permissions}:
$\RW$, $\RO$, $\WO$, and $\NO$, which form a diamond lattice:
$\RW > \RO | \WO > \NO$. We denote permissions with $\pi$ and
access policies with $\Pi$.

\newcommand{\fresh}[1]{\mathrm{fresh} \: #1}
\newcommand{\test}[1]{\mathrm{check} \: #1}
\newcommand{\freeze}{\mathrm{freeze}}
\newcommand{\lift}{\mathrm{lift}}
\newcommand{\block}{\mathrm{block}}
\newcommand{\drop}{\mathrm{drop}}
\newcommand{\cut}{\mathrm{cut}}
\newcommand{\move}{\mathrm{move}}
\newcommand{\observe}{\mathrm{observe}}
\newcommand{\borrow}{\mathrm{borrow}}

Permission transformers modify policies at a given path,
as well as its prefixes and extensions.
Symbolically, we write $\Pi \xrightarrow{T}_p \Pi'$
to denote that policy $\Pi'$ results from application of
transformer $T$ to $\Pi$ at path $p$.
We write $ \Pi \xrightarrow{T_1 \comp\, T_2}_p \Pi'$
as an abbreviation for
$\Pi \xrightarrow{T_1}_p \comp \xrightarrow{T_2}_p \Pi'$
(that is, for some $\Pi''$,
$\Pi \xrightarrow{T_1}_p \Pi'' \xrightarrow{T_2}_p \Pi'$).
We write $ \Pi \xrightarrow{T}_{p,q} \Pi' $
as an abbreviation for
$\Pi \xrightarrow{T}_p \comp \xrightarrow{T}_q \Pi'$.

Permission transformers can also apply to
expressions, which consists in updating the policy
for every path in the expression.
This only includes paths that occur as sub-expressions:
in an expression \texttt{X.f.g + Y.h}, only the paths
\texttt{X.f.g} and \texttt{Y.h} are concerned, whereas
\texttt{X}, \texttt{X.f} and \texttt{Y} are not.
The order in which the individual paths are treated
must not affect the final result.

\begin{figure}[t]

\infrule[P-assign]{
\Pi \xrightarrow{\move}_e \comp
\xrightarrow{\test{\WO}\: \comp \:\fresh{\RW}\: \comp \:\lift}_p \Pi'}
{\Pi \cdot p \,\texttt{:=}\, e \rightarrow \Pi'}

\medskip
\infrule[P-alloc]{
\Pi \xrightarrow{\test{\WO}}_p \comp
    \xrightarrow{\fresh{\WO}\: \comp \:\cut\: \comp \:\block}_{p.\texttt{all}} \Pi'}
{\Pi \cdot p \,\texttt{:=}\, \texttt{new}\:\:\tau \rightarrow \Pi'}

\medskip
\infrule[P-if]{
\Pi \xrightarrow{\test{\RO}}_e \Pi \andalso
\Pi \cdot \bar{s}_1 \rightarrow \Pi_1 \andalso
\Pi \cdot \bar{s}_2 \rightarrow \Pi_2 \andalso
\forall p.\, \Pi'(p) = \Pi_1(p) \land \Pi_2(p)}
{\Pi \cdot \texttt{if}\; e \,\:\texttt{then}\; \bar{s}_1
\;\texttt{else}\; \bar{s}_2 \;\texttt{end} \rightarrow \Pi'}

\medskip
\infrule[P-while]{
\Pi \xrightarrow{\test{\RO}}_e \Pi \andalso
\Pi \cdot \bar{s} \rightarrow \Pi' \andalso
\forall \pi.\, \Pi'(\pi) \geqslant \Pi(\pi)}
{\Pi \cdot \texttt{while}\; e \,\:\texttt{loop}\; \bar{s}
\;\texttt{end} \rightarrow \Pi}

\medskip\smallskip
\infrule[P-call]{
\texttt{procedure} \:P\:
\texttt{(}\,
a_1 \,\texttt{:}\,\texttt{in}\: \tau_{a_1} \texttt{;} \ldots \texttt{;}\,
b_1 \,\texttt{:}\,\texttt{in-out}\: \tau_{b_1} \texttt{;} \ldots \texttt{;}\,
c_1 \,\texttt{:}\,\texttt{out}\: \tau_{c_1} \texttt{;} \ldots \texttt{)} \\
\text{~~~~~~~~~~~~~~~~~~~~~~~~~~~~}
\texttt{is}\; \cdots \:
\texttt{begin} \; \bar{s} \; \texttt{end}
\text{~~is declared in the program}
\\[1ex]
\Pi \xrightarrow{\test{\RO} \:\comp\: \observe}_{e_{a_1}, \ldots}
\comp \xrightarrow{\test{\RW} \:\comp\: \borrow}_{p_{b_1},\ldots}
\comp \xrightarrow{\test{\WO} \:\comp\: \borrow}_{q_{c_1}, \ldots} \Pi'' \\
\Pi \xrightarrow{
  \fresh{\RW}\: \comp \:\lift}_{p_{b_1},\ldots, q_{c_1}, \ldots} \Pi'
}
{\Pi \cdot
P(e_{a_1},\ldots, p_{b_1},\ldots,q_{c_1},\ldots) \rightarrow \Pi'}
\caption{Alias safety rules for statements.}\label{fig:rules}
\end{figure}

We define the rules of alias safety for {\muSPARK}
statements in the context of a current access policy.
An \emph{alias-safe} statement yields an updated policy
which is used to check the subsequent statement.
We write $\Pi \cdot s \rightarrow \Pi'$ to denote that statement $s$
is safe with respect to policy $\Pi$ and yields the updated
policy $\Pi'$. We extend this notation to sequences of statements
$\bar{s}$ in an obvious way. The rules for checking the alias safety
of statements are given in Fig.~\ref{fig:rules}.
These rules use a number of permission transformers such
as `fresh', `check', `move', `observe', and `borrow', which
we define and explain below.

Let us start with the (\textsc{P-assign}) rule.
Assignments grant the full ownership over the copied value
to the left-hand side. If we copy a value of a shallow type,
we merely have to
ensure that the right-hand side has the read permission.
Whenever we copy a deep-typed value, aliases may be created,
and we must check that the right-hand side is initially
the sole owner of the copied value (that is, possesses the
$\RW$ permission) and revoke the ownership from it.

To define the `${\move}$' transformer that handles
permissions for the right-hand side of an assignment,
we need to introduce several simpler transformers.

\begin{definition}
Permission transformer\/ $\test{~\pi}$ does not modify the
access policy and only verifies that a given path $p$
has permission $\pi$ or greater. In other words,
$\Pi \xrightarrow{\test{\pi}}_p \Pi'$ if and only if
$\Pi(p) \geqslant \pi$ and $\Pi = \Pi'$.
This transformer also applies to expressions:
$\Pi \xrightarrow{\test{\pi}}_e \Pi'$ states
that $\Pi \xrightarrow{\test{\pi}}_p \Pi' (= \Pi)$
for every path $p$ occurring in $e$.
\end{definition}

\begin{definition}
Permission transformer $\fresh{\pi}$ assigns permission $\pi$
to a given path $p$ and all its extensions.
\end{definition}

\begin{definition}
Permission transformer\/ $\cut$ assigns restricted permissions
to a deep path $p$ and its extensions:
the path $p$ and its near deep extensions receive
permission $\WO$,
the near shallow extensions keep their current permissions,
and the far extensions receive permission $\NO$.
\label{def:cut}
\end{definition}

Going back to the procedure \texttt{P1} in Fig.~\ref{fig:P1P2},
the change of permissions on the right-hand side after the
assignment \texttt{A} \texttt{:=} \texttt{B} corresponds to
the definition of `$\cut$'.
In the case where the right-hand side of an assignment is
not simply a variable, but a deep path or a \texttt{'Access}
expression, we also need to change permissions of the prefixes,
to reflect the ownership transfer.

\begin{definition}
Permission transformer\/ $\block$ propagates the loss of
the read permission from a given path to all its prefixes.
Formally, it is defined by the following rules, where $x$
stands for a variable and $f$ for a field name:
\[\begin{array}{ccc}
\dfrac{}{\Pi \xrightarrow{\block}_x \Pi}
&&
\dfrac
{ \Pi[p \mapsto \WO]\! \xrightarrow{\block}_p \!\Pi'}
{\Pi \xrightarrow{\block}_{p.\textnormal{\texttt{all}}} \Pi'}
\\[4ex]
\dfrac
{ \Pi(p) = \NO }
{\Pi \xrightarrow{\block}_{p.f} \Pi}
&\qquad&
\dfrac
{ \Pi(p) \geqslant \WO  \quad
      \Pi[p \mapsto \WO]\! \xrightarrow{\block}_p \!\Pi'}
{\Pi \xrightarrow{\block}_{p.f} \Pi'}
\end{array}\]
\end{definition}

\begin{definition}
Permission transformer\/ $\drop$ propagates the loss of both
read and write permissions from a given path to its prefixes
up to the first pointer, and propagates the loss of the
read permission afterwards:
\[\begin{array}{c}
\dfrac{}{\Pi \xrightarrow{\drop}_x \Pi}
\qquad\quad
\dfrac
{ \Pi[p \mapsto \WO]\! \xrightarrow{\block}_p \!\Pi'}
{\Pi \xrightarrow{\drop}_{p.\textnormal{\texttt{all}}} \Pi'}
\qquad\quad
\dfrac
{ \Pi[p \mapsto \NO]\! \xrightarrow{\drop}_p \!\Pi'}
{\Pi \xrightarrow{\drop}_{p.f} \Pi'}
\end{array}\]
\end{definition}


\begin{definition}
Permission transformer $\move$ applies to expressions:
\vspace{-\topsep}
\begin{itemize}
\item if $e$ has a shallow type, then
$\Pi \xrightarrow{\move}_e \Pi' \,\Leftrightarrow\,
\Pi \xrightarrow{\test{\RO}}_e \Pi'$;
\item if $e$ is a deep path $p$, then
$\Pi \xrightarrow{\move}_e \Pi' \,\Leftrightarrow\,
\Pi \xrightarrow{\test{\RW}\: \comp \:\cut\: \comp \:\block}_p \Pi'$;
\item if $e$ is $p\texttt{'Access}$, then
$\Pi \xrightarrow{\move}_e \Pi' \,\Leftrightarrow\,
\Pi \xrightarrow{\test{\RW}\: \comp \:\fresh{\NO}\: \comp \:\drop}_p \Pi'$;
\item if $e$ is \texttt{null}, then
$\Pi \xrightarrow{\move}_e \Pi' \,\Leftrightarrow\, \Pi' = \Pi$.
\end{itemize}
\end{definition}

To further illustrate the `$\move$' transformer, let us consider
two variables \texttt{P} and \texttt{Q} of type \texttt{access List}
and an assignment \texttt{P := Q.all.Next}. We assume that
\texttt{Q} and all its extensions have full ownership ($\RW$)
before the assignment.
We apply the second case in the definition of `$\move$' to
the deep path \texttt{Q.all.Next}. The `$\test{\RW}$' condition
is verified, and the `$\cut$' transformer sets the permission
for \texttt{Q.all.Next} to $\WO$ and the permission for
\texttt{Q.all.Next.all} and all its extensions to $\NO$.
Indeed, \texttt{P.all} becomes an alias of \texttt{Q.all.Next.all}
and steals the full ownership for this memory area. However,
we still can reassign \texttt{Q.all.Next} to a different
address. Moreover, we still can write some new values into
\texttt{Q.all} or \texttt{Q}, without compromising safety.
This is enforced by the application of the `$\block$' transformer
at the end. We cannot keep the read permission for
\texttt{Q} or \texttt{Q.all}, since it implies the read access
to the data under \texttt{Q.all.Next.all}.

Now, let a variable \texttt{R} have type
\texttt{access} \texttt{Boolean} and consider the assignment
\texttt{R := Q.all.Flag'Access}. We apply the third case
in the definition of `$\move$'. Assuming once again that
\texttt{Q} has full ownership over its value, the
`$\test{\RW}$' condition for \texttt{Q.all.Flag}
is verified. Since the ownership of this Boolean value
is now transferred to \texttt{R.all}, we must revoke
all permissions from \texttt{Q.all.Flag}
(and its extensions, if it had any), which is enforced
by `$\fresh{\NO}$'. Moreover, since writing into
the record \texttt{Q.all} overwrites the \texttt{Flag}
field, we must also revoke all permissions from \texttt{Q.all}.
This is done by the `$\drop$' transformer. Notice that the
permissions for \texttt{Q.all.Key} and \texttt{Q.all.Next}
are not affected: we can still read and modify those
fields, as they are not aliased with other paths. Furthermore,
modifying the pointer \texttt{Q} itself is allowed, which is
why `$\drop$' becomes `$\block$' after rising past \texttt{all}.

Finally, we need to describe the change of permissions on
the left-hand side of an assignment, in order to reflect
the gain of the full ownership.
The idea is that as soon as we have the full ownership
for each field of a record, we can assume the full ownership
of the whole record, and similarly for pointers.
\begin{definition}
Permission transformer\/ $\lift$ propagates the $\RW$
permission from a given path to its prefixes, wherever
possible:
\[\begin{array}{ccc}
\dfrac{}{\Pi \xrightarrow{\lift}_x \Pi}
&
\qquad
&
\dfrac
{ \Pi[p \mapsto \RW]\! \xrightarrow{\lift}_p \!\Pi'}
{\Pi \xrightarrow{\lift}_{p.\textnormal{\texttt{all}}} \Pi'}
\\[4ex]
\dfrac
{ \forall q \sqsupset p.\, \Pi(q) = \RW \quad
  \Pi[p \mapsto \RW]\! \xrightarrow{\lift}_p \!\Pi'}
{\Pi \xrightarrow{\lift}_{p.f} \Pi'}
&&
\dfrac
{ \exists q \sqsupset p.\, \Pi(q) \neq \RW }
{\Pi \xrightarrow{\lift}_{p.f} \Pi}
\end{array}\]
\end{definition}

In the (\textsc{P-assign}) rule, we revoke the permissions
from the right-hand side of an assignment before granting
the ownership to the left-hand side. This is done in order
to prevent creation of circular data structures. Consider an assignment
\texttt{A.Next := A'Access}, where \texttt{A} has type \texttt{List}.
According to the definition of `$\move$', path \texttt{A} and all
its extensions receive permission $\NO$. This makes the left-hand
side \texttt{A.Next} fail the write permission check.

Allocations \texttt{p := new $\tau$} are handled by the
(\textsc{P-alloc}) rule. As long as
the memory area under the pointer \texttt{p} is not explicitly
initialised by the program code, no read permission is granted
for \texttt{p}, nor its extensions and prefixes. Moreover, since
the pointer fields (if any) in the allocated memory are not
accessible yet, no permission at all can be given for the far
extensions of \texttt{p}. This is enforced by the `$\cut$'
transformer.

In a conditional statement,
the policies at the end of the two branches are merged
selecting the most restrictive permission for each path.
Loops require that no permissions are lost at the end of
a loop iteration, compared to the entry, as explained above
for procedure~\texttt{P2} in Fig.~\ref{fig:P1P2}.

Procedure calls guarantee to the callee that every argument with mode
\texttt{in}, \texttt{in-out}, or \texttt{out} has at least permission
$\RO$, $\RW$ or $\WO$, respectively. To ensure the absence of potentially
harmful aliasing, we revoke the necessary permissions using
the `$\observe$' and `$\borrow$' transformers.

\begin{definition}
Permission transformer\/ $\borrow$ assigns permission $\NO$
to a given path $p$ and all its prefixes and extensions.
\end{definition}

\begin{definition}
Permission transformer\/ $\freeze$ removes the write permission
from a given path $p$ and all its prefixes and extensions.
In other words, $\freeze$ assigns to each path $q$
comparable to $p$ the minimum permission $\Pi(q) \land \RO$.
\end{definition}

\begin{definition}
Permission transformer\/ $\observe$ applies to expressions:
\vspace{-\topsep}
\begin{itemize}
\item if $e$ has a shallow type, then
$\Pi \xrightarrow{\observe}_e \Pi' \,\Leftrightarrow\,
\Pi' = \Pi$;
\item if $e$ is a deep path $p$, then
$\Pi \xrightarrow{\observe}_e \Pi' \,\Leftrightarrow\,
\Pi \xrightarrow{\freeze}_p \Pi'$;
\item if $e$ is $p\texttt{'Access}$, then
$\Pi \xrightarrow{\observe}_e \Pi' \,\Leftrightarrow\,
\Pi \xrightarrow{\freeze}_p \Pi'$;
\item if $e$ is \texttt{null}, then
$\Pi \xrightarrow{\observe}_e \Pi' \,\Leftrightarrow\, \Pi' = \Pi$.
\end{itemize}
\end{definition}

We remove the write permission from the deep-typed
\texttt{in} parameters using the `$\observe$' transformer,
in order to allow aliasing between the read-only paths.
As for the \texttt{in-out} and \texttt{out} parameters,
we transfer the full ownership over them to the callee,
which is reflected by dropping every permission on the
caller's side using `$\borrow$'.

In the (\textsc{P-call}) rule, we revoke permissions
right after checking them for each parameter. In this way,
we cannot pass, for example, the same path as an \texttt{in}
and \texttt{in-out} parameter in the same call. Indeed,
the `$\observe$' transformer will remove the write permission,
which is required by `$\test{\RW}$' later in the transformer chain.
At the end of the call, the callee transfers to the caller the
full ownership over each \texttt{in-out} and \texttt{out} parameter.

We apply our alias safety analysis to each procedure declaration.
We start with an empty access policy, denoted $\varnothing$.
Then we fill the policy with the permissions for the formal
parameters and the local variables and check the procedure body.
At the end, we verify that every \texttt{in-out} and \texttt{out}
parameter has the $\RW$ permission. Formally, this is expressed
with the following rule: 
\infrule
{\varnothing \xrightarrow{\fresh{\RO}}_{a_1,\ldots} \comp
 \xrightarrow{\fresh{\RW}}_{b_1,\ldots} \comp
 \xrightarrow{\fresh{\WO} \:\comp\: \cut}_{c_1,\ldots} \comp
 \xrightarrow{\fresh{\WO} \:\comp\: \cut}_{d_1,\ldots} \Pi' \\[0.7ex]
 \Pi' \cdot \bar{s} \rightarrow \Pi'' \andalso
 \Pi''(b_1) = \cdots = \Pi''(c_1) = \cdots = \RW}
{\texttt{procedure} \:P\:
  \texttt{(}\,
  a_1 \,\texttt{:}\,\texttt{in}\: \tau_{a_1} \texttt{;} \ldots \texttt{;}\,
  b_1 \,\texttt{:}\,\texttt{in-out}\: \tau_{b_1} \texttt{;} \ldots \texttt{;}\,
  c_1 \,\texttt{:}\,\texttt{out}\: \tau_{c_1} \texttt{;} \ldots \texttt{)} \\
\text{~~~~~~~~~~~~~~~~~~~~~~~~~~~~~~~~~~~~}
\texttt{is}\; d_1 \,\texttt{:}\, \tau_{d_1} \texttt{;} \ldots \:
\texttt{begin} \; \bar{s} \; \texttt{end} \text{~~~is alias-safe}}
We say that a {\muSPARK} program is \emph{alias-safe}
if all its procedures are.

\subsection{Soundness}

As the end of the analysis, an alias-safe program has
an access policy associated to each sequence point in it.
We say that an access policy $\Pi$ is \emph{consistent}
whenever it satisfies the following conditions for all
valid paths $\pi$, $\pi.f$, $\pi.\texttt{all}$:
\begin{align}
\Pi(\pi) = \RW &\Longrightarrow \Pi(\pi.f) = \RW &
\Pi(\pi) = \RW &\Longrightarrow \Pi(\pi.\textnormal{\texttt{all}}) = \RW \\
\Pi(\pi) = \RO &\Longrightarrow \Pi(\pi.f) = \RO &
\Pi(\pi) = \RO &\Longrightarrow \Pi(\pi.\textnormal{\texttt{all}}) = \RO \\
\Pi(\pi) = \WO &\Longrightarrow \Pi(\pi.f) \geq \WO
\end{align}
These invariants correspond to the informal explanations
given in Section~\ref{sec:pointer-analysis}. Invariant (1)
states that the full ownership over a value propagates
to all values reachable from it. Invariant (2) states
that the read-only permission must also propagate to
all extensions. Indeed, a modification of a reachable component
can be observed from any prefix. Invariant (3) states
that write permission over a record value implies a write
permission over each of its fields. However, the write
permission does not necessarily propagate across pointer
dereference.

\begin{lemma}[Policy consistency]
The alias safety rules in Fig.~\ref{fig:rules}
preserve policy consistency.
\end{lemma}

When, during an execution, we arrive at a given sequence point
with the set of variable bindings $\Upsilon$, store $\Sigma$,
and statically computed and consistent access policy $\Pi$,
we say that the state of the execution respects the
\textsl{Concurrent Read, Exclusive Write} condition (CREW),
if and only if for any two distinct valid paths $p$ and $q$,
$\lval{p}^\Upsilon_\Sigma = \lval{q}^\Upsilon_\Sigma \land
\Pi(p) \geq \WO \Longrightarrow \Pi(q) = \NO$.

The main result about the soundness of our approach
is as follows:
\begin{theorem}[Soundness]
A terminating evaluation of a well-typed alias-safe {\muSPARK} program
respects the CREW condition at every sequence point.
\end{theorem}

The full proof, for a slightly different definition of {\muSPARK},
is given in~\cite{arxiv}. The argument proceeds by induction on the
evaluation derivation, following the rules in Figure~\ref{fig:sem}.
The only difficult cases are assignment, where the required
permission withdrawal is ensured by the `$\move$' transformer,
and procedure call, where the chain of `$\observe$' and
`$\borrow$' transformers, together with the corresponding checks,
on the caller's side, ensures that the CREW condition is respected
at the beginning of the callee.

\section{Implementation}
\label{sec:experimental}
\label{sec:completeSPARK}

We have implemented the permission rules in the GNAT compiler for
Ada, which is part of GCC. The implementation
consists in 3700 lines in Ada. The analysis is
triggered by using the debug switch \texttt{-gnatdF} and setting the context
for formal verification of SPARK code with switch \texttt{-gnatd.F} and
aspect \texttt{SPARK\_Mode}.

Access policies are infinitely big in presence of recursive types,
which we took into account in our implementation with a lazy
implementation of permission trees. Permission trees start with
a depth of one, and are expanded on demand.

The permission rules presented in Section~\ref{sec:pointer-analysis} only
address a subset of SPARK. Complete SPARK differs from {\muSPARK} on several
points, which have been taken into account when designing the full set of
rules. For arrays, permission rules are adapted to apply to all elements,
without taking into account the exact index of that element, which may not be
known statically in general. Besides procedures, SPARK has functions, which
return values and cannot perform side-effects. Functions take only parameters
of mode \texttt{in} and can be called inside expressions. Our permission rules
are augmented with an implicit $\move$ of the values returned by functions,
which allows us to support constructors.

In our formalization, we considered that every shallow
\texttt{in} parameter is passed by-copy, which is not the case in SPARK. In our
implementation, we correctly distinguish parameters of by-copy types (typically
scalars) and parameters which may be passed by-copy or by-reference, which we
treat as deep parameters (except for function \texttt{in} parameters,
as functions cannot have side effects, and in particular cannot write in
their parameters even under dereference).

Loops in SPARK also differ on two accounts from the
loops in our formalization: besides `while' loops, SPARK also defines `for' loops
and plain loops (with no exit condition), and exit statements inside a loop allow
exiting any enclosing loop. In our implementation, we take these into account to
create the correct merged context for analyzing the code after the
loop.


The analysis has been tested on four test suites,
including the regression testsuite of the GNAT compiler (17041 tests \cite{ACATS})
and the regression testsuite of the SPARK product (2087 tests), which
detected 30 regressions, most of them related to object oriented features
that the current analysis is not able to handle.
We also wrote a dedicated testsuite \cite{githubtestsuite}
containing 20 tests inspired by the examples given in
the Rust borrow-checker documentation~\cite{bckRust}.

\section{Related Works}
\label{sec:related}

    \subsection{Comparison with Rust}

The following section compares Rust and SPARK on some features and
constructs that seem relevant to the authors. Both prevent harmful
aliasing in the source code, whereas different design choices affect
their respective expressiveness. In SPARK, an additional limitation
was the choice not to add any annotation or keyword to the language.

The main differences come from the fact that SPARK uses other
compiler passes to handle many safety features, whereas they are
handled directly by Rust's borrow-checker and type-checker.
A benefit of our work is that it unambiguously defines the rules
for SPARK while there exists no official document specifying Rust's
borrow-checker, in all details, especially as Rust continues to go
through significant evolutions, such as non-lexical lifetimes~\cite{RFC2094}.
We must note, however, a significant recent effort to provide a rigorous
formal description of the foundations of Rust~\cite{rustbelt}.

In SPARK, the duration of borrows is limited to the duration of procedure calls.
The mechanism of \emph{renaming} in SPARK can be used to shorten long paths, and
thus it is less important to be able to create several local copies of the
same deep variable.
It turns out, however, that this restriction forbids traversing a linked data
structure with only $\RO$ permission, and that even with $\RW$ permission the
structure needs to be reconstructed after traversal. We are working on the
rules for local application of borrow and observe inside a block
to allow these.

Rust has sophisticated lifetime checks, allowing to precisely control
the duration of a borrowed pointer. In SPARK, similar checks are
implemented as a separate analysis pass of GNAT compiler, with their
own set of rules that are less expressive: the lifetime of a pointer
is limited to the block in which its type has been declared.

There are no null pointers in Rust, whereas they are allowed in SPARK.
Dereferencing a null pointer is a runtime error, and the null-pointer
safety must be proved by a separate analysis by the SPARK verification
tools~\cite{thalesada}.

Another difference is that mutable borrows in Rust guarantee the
ownership of the underlying memory at any time, whereas in SPARK
borrow guarantees ownership only at the entry and exit of the procedure.
In particular, it is not possible in Rust to move out a borrowed
variable without assigning it in the same statement. In SPARK,
any borrowed variable can be moved as long as it is assigned
before the end of the procedure, which grants it the $\RW$
permission. This allows us to directly implement the swap
procedure in SPARK, whereas the Rust implementation relies
on unsafe operations:
\begin{lstlisting}[style=spark]
  procedure Swap (X, Y : in out T) is
    Temp : T := Y;  -- Move Y. X:RW, Y:W, Temp:RW
  begin
    Y := X;         -- Move X. X:W, Y:RW, Temp:RW
    X := Temp;      -- Move Temp. X:RW, Y:RW, Temp:W
  end Swap;         -- Both arguments X and Y are RW.
\end{lstlisting}

Finally, it is impossible to implement cycling constructs in
SPARK (the right-hand side of an assignment cannot be an ancestor
of the left-hand side). Rust has a similar limitation and requires
some workarounds like reference counting to implement structures
like graphs~\cite{RustGraphs}. Nevertheless, both Rust and SPARK
allow compiling with parts of code written in an unsafe superset
of the language. This is the case for some standard library
containers (hash tables, iterators, smart pointers, etc.) that
have an interface specified in SPARK and Rust, and can be called
safely from safe Rust/SPARK.

    \subsection{Other Related Works}

Permission-based programming languages generalize the issue of avoiding
harmful aliasing to the more general problem of preventing harmful sharing of
resources (memory, but also network connections, files, etc.). Cyclone and Rust
achieve absence of harmful aliasing by enforcing an ownership type system on
the memory pointed to by objects~\cite{Grossman2002,Balasubramanian}.

Dafny associates each object with its \emph{dynamic frame}, the set
of pointers that it owns~\cite{Leino2010}. This dynamic version of
ownership is enforced by modeling the ownership of pointers in logic,
generating verification conditions to detect violations of the
single-owner model, and proving them using SMT provers. In Spec\#,
ownership is similarly enforced by proof, to detect violations of
the so-called Boogie methodology~\cite{Boogie}.

Separation logic~\cite{seplogic} is an extension of Hoare-Floyd logic
that allows reasoning about pointers. In general, it is difficult
to integrate into automated deductive verification: in particular,
it is not directly supported by SMT provers.

In our work, we use a permission-based mechanism for detecting potentially
harmful aliasing, in order to make the presence of pointers transparent for
automated provers. In addition, our approach does not require additional user
annotations, that are required in some of the previously mentioned techniques.
We thus expect to achieve high automation and usability, which was our goal
for supporting pointers in SPARK.

    \section{Conclusion}

In this paper, we have presented anti-aliasing rules that allow
supporting pointers in SPARK. We showed a systematic analysis
that allows a wide range of use cases with pointers and dynamic
allocation. To the best of our knowledge, this is a novel approach
for controlling aliasing introduced by arbitrary pointers in
a programming language supported by proof. Our approach does not
require user annotations or proof of verification conditions,
which makes it much simpler to adopt. Moreover, we provided
a formalization of our rules on a subset of SPARK in order
to mathematically prove the safety of our analysis.
Finally, we compared our method to the Rust language which
provides a similar analysis.

More work needs to be done to fully support pointers in SPARK.
Both flow analysis and proof need to be adapted to account for
the presence of pointers. This work has started and is expected
to be completed by the end of 2018. This will make it possible
to use formal verification with SPARK on industrial programs
with pointers, something that was long believed to be impossible.

We also need to extend our formalism and proof to non-terminating
executions. For that purpose, we can provide a co-inductive
definition of the big-step semantics and perform a similar
co-inductive soundness proof, as described by Leroy and
Grall~\cite{Leroy_2009}.

The formal study of the constructs that can be implemented using
this borrow-checker has not been discussed in this paper. Such
a study could allow a formalization of anti-aliasing analyses
through classes of expressiveness. Another long-term goal would
be extending our analysis so that it could handle automatic
reclamation, parallelism, initialization and lifetime checks,
instead of relying on external checks.

The proposed anti-aliasing rules are being discussed by the
Ada Rapporteur Group for inclusion in the next version of
Ada~\cite{AI2018}.


\clearpage
\bibliographystyle{splncs}
\bibliography{report}
\end{document}

%% file: macros.tex
\usepackage{xcolor}
\definecolor{commentgreen}{RGB}{34,140,34}
\definecolor{keywordblue}{RGB}{0,0,255}
\definecolor{stringpurple}{RGB}{161,32,241}

\lstdefinestyle{spark}{
	language=Ada,
	keywordstyle=\bfseries\ttfamily\color{keywordblue},
	identifierstyle=\ttfamily,
	commentstyle=\ttfamily\color{commentgreen},
	stringstyle=\ttfamily\itshape\color{stringpurple},
	morekeywords=[1]some,
	showstringspaces=false,
	basicstyle=\small\ttfamily,
	stepnumber=1,
	numbersep=10pt,
	tabsize=2,
	breaklines=true,
	prebreak = \raisebox{0ex}[0ex][0ex]{\ensuremath{\hookleftarrow}},
	breakatwhitespace=false,
	columns=fixed,
	extendedchars=true,
        literate={>=}{>{}={}}{2}{<=}{<{}={}}{2}
}

\lstdefinelanguage{Rust}{%
  basicstyle=\small\ttfamily,
  columns=[l]fullflexible,sensitive=true,keepspaces=true,%
  morekeywords=[1]{let,mut},%
  keywordstyle=[1]{\bfseries\ttfamily\color{keywordblue}},%
  morecomment=[l]{//},%
  commentstyle={\ttfamily\color{commentgreen}},%
}

